# Angular distribution of radiation by relativistic electrons in a thin crystal

X. Artru [a], S.P. Fomin [b,*], N.F. Shul'ga [b,c]

[a] *Lyon Institute of Nuclear Physics, Villeurbanne 69622 Cedex, France*
[b] *Kharkov Institute of Physics and Technology, Kharkov 61108, Ukraine*
[c] *Belgorod State University, Belgorod 307007, Russia*

**Abstract**

The results of theoretical investigation of angular distributions of radiation from a relativistic electron passing through a thin crystal at a small angle to the crystal axis are presented. The electron trajectories in crystal were simulated using the binary collision model which takes into account both coherent and incoherent effects at scattering. The angular distribution of radiation was calculated as a sum of radiation from each electron. It is shown that there are nontrivial angular distributions of the emitted photons, which is connected to the superposition of the coherent scattering of electrons by atomic rows ("doughnut scattering" effect) and the suppression of the radiation due to the multiple scattering effect (similar to the Landau-Pomeranchuk-Migdal effect in an amorphous matter). The orientation dependence of angular distribution of radiation is also analyzed.

## 1. Introduction

The multiple scattering of a high energy electron on atoms in an amorphous matter can break the dipole regime of radiation and, as a result, can suppress the radiation of relatively soft photons. This effect is known as the Landau-Pomeranchuk-Migdal effect (the LPM effect) [1]. It appears when the mean square angle of the electron multiple scattering $\overline{\vartheta}_e$ during one coherence length of bremsstrahlung [2] becomes larger than the characteristic angle $\theta \sim \gamma^{-1}$ of radiation of a relativistic particle, where $\gamma$ is the Lorenz-factor of the particle. A similar effect takes place when relativistic electrons pass through a crystal under a small angle $\psi$ to a crystal axis [3]. In this case there is a suppression of coherent bremsstrahlung. In [4] it was noted that the condition of the LPM effect in crystal can be fulfilled at a smaller energy of the projectile particles than in an amorphous matter, because the multiple scattering in a crystal has a coherent character ("doughnut scattering" effect) and the root-mean-square angle of the multiple scattering $\bar{\vartheta}_{cr}$ can exceed significantly the corresponding value for an amorphous target of the same thickness [5]. A crystal is also suitable for study the LPM effect due to the additional possibility to change the efficiency of multiple scattering by changing the orientation angle $\psi$ instead of the thickness changing as in the amorphous target case.

[*] Corresponding author. Tel.: +380 572 352 115; fax: +380 572 352 683.
 *E-mail address*: sfomin@kipt.kharkov.ua

Theoretical and experimental investigations of the LPM effect usually consider the influence of multiple scattering on the spectral density of radiation (see, for example, the review [6] and references therein). However, as it was shown in [7] by the example of radiation in a thin amorphous target, the multiple scattering could change essentially not only the spectrum of the emitted photons [8] but also the form of their angular distribution.

The present work is devoted to a theoretical investigation of the angular distribution of radiation by relativistic electrons in a thin crystal under the LPM effect condition. The electron trajectories in the crystal were simulated using the binary collision model, which takes into account both coherent and incoherent scattering.

The calculated angular distribution of radiation shows a peculiar peak structure, which changes considerably as the incidence angle of the electrons varies. This structure is attributed to the superposition of the coherent scattering of the electrons by atomic strings ("doughnut scattering") and the non-dipole regime of radiation (as it is for the Landau-Pomeranchuk-Migdal effect in an amorphous target), when the root-mean-square value of the multiple scattering angle becomes larger than characteristic radiation angle $\theta \sim \gamma^{-1}$.

## 2. General formulas

The spectral-angular density of radiation by an electron of trajectory $\vec{r}(t)$ is determined in classical electrodynamics by the expression [9,10]

$$\frac{d^2E}{d\omega do} = \frac{e^2}{4\pi^2}\left[\vec{k}\times\vec{I}\right]^2, \qquad (1)$$

where $\vec{k}$ and $\omega$ are the wave vector and the frequency of the radiated wave and

$$\vec{I} = \int_{-\infty}^{\infty} \vec{v}(t)\, e^{i(\omega t - \vec{k}\vec{r})} dt. \qquad (2)$$

In a thin layer of matter the characteristic values $\vartheta_e$ of the scattering angles of a relativistic electron are small in comparison with unity. If the coherence length of radiation process is big in comparison with the thickness of the target

$$l_c \approx \frac{2\gamma^2}{\omega} \frac{1}{1+\gamma^2\theta^2+\gamma^2\vartheta_e^2} \gg T, \quad (3)$$

then $\vec{I}$ can be represented as [7]

$$\vec{I} \approx \frac{i}{\omega}\left(\frac{\vec{v}'}{1-\vec{n}\vec{v}'} - \frac{\vec{v}}{1-\vec{n}\vec{v}}\right), \quad (4)$$

where $\vec{v}$ and $\vec{v}'$ are the electron velocities before and after scattering, $\vec{n} = \vec{k}/\omega$.

The spectral-angular density of the radiation in this case is determined only by the scattering angle of the particle in matter. Putting (4) into (1), we obtain

$$\frac{d^2 E}{d\omega\, do} = \frac{e^2\gamma^2}{\pi^2 p}\left[\frac{1+\alpha^2+\alpha^2\beta^2+2\alpha\beta\cos\varphi}{(1+\alpha^2)^2} - \frac{1}{p}\right], \quad (5)$$

where $p = 1+\alpha^2+\beta^2-2\alpha\beta\cos\varphi$, $\alpha = \gamma\theta$, $\beta = \gamma\vartheta_e$, $\theta$ and $\varphi$ are the polar and azimuthal angles of radiation. The angles $\theta$ and $\vartheta_e$ are counted from the direction of the initial velocity $\vec{v}$ of the electron. $\varphi$ is the angle between the vectors $\vec{k}_\perp$ and $\vec{v}'_\perp$ in the plane orthogonal to $\vec{v}$.

Note that formula (5) does not contain the photon frequency dependence immediately. However, the validity of condition (3) for this formula depends on it. So, all the following results are valid for relatively soft gamma-quanta defined by inequality (3). For example, for 10 GeV electrons this condition is ω < 15 MeV. It should be also noted that the region of validity of formula (5) widens rapidly (~ $\gamma^2$) as the electron energy grows.

Looking at some features of the angular distributions of relativistic electrons radiation in a thin layer of matter, let's first consider the angular distribution of electron radiation in the plane $(\vec{v}, \vec{v}')$. Turning in (5) to the Cartesian coordinates $(\alpha_x = \alpha\cos\varphi, \alpha_y = \alpha\sin\varphi)$, we find, that for $\alpha_y = 0$

$$\frac{d^2 E}{d\omega\, do} = \frac{e^2\gamma^2}{\pi^2}\left[\frac{\alpha_x}{1+\alpha_x^2} - \frac{\alpha_x - \beta}{1+(\alpha_x - \beta)^2}\right]^2. \quad (6)$$

This result is illustrated in Fig. 1, which displays the angular distributions of radiation by the electron in the $(\mathbf{v}, \mathbf{v}')$ plane for different values of the parameter $\beta = \gamma\vartheta_e$.

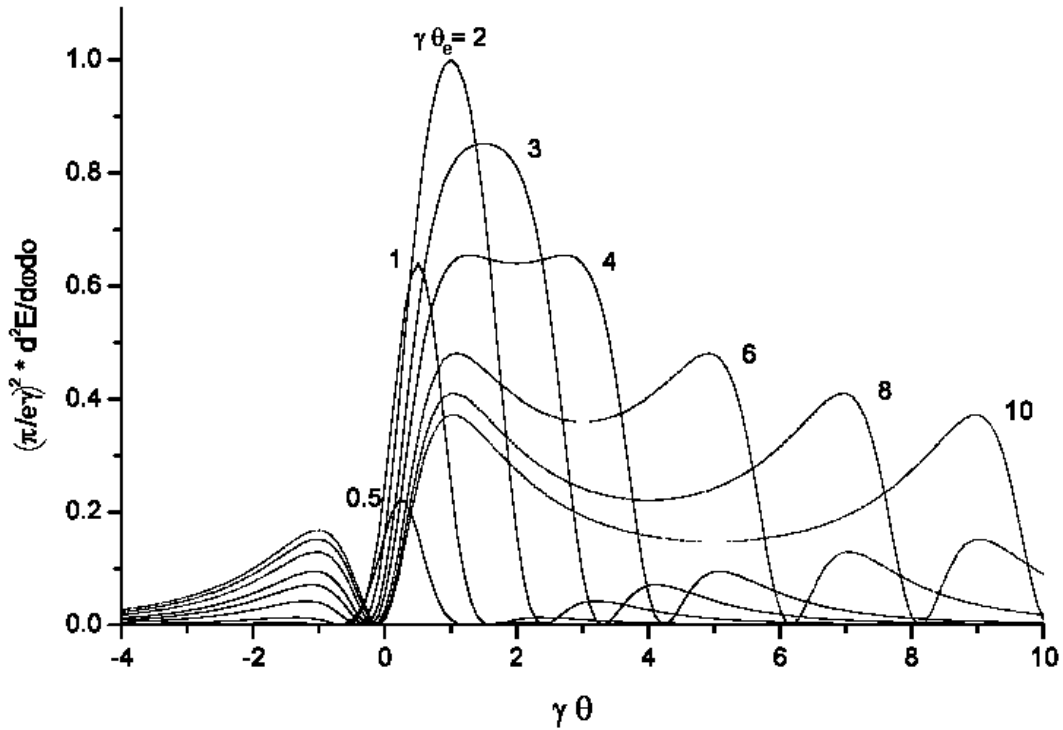

Fig. 1. The angular distributions of radiation in the $(\mathbf{v}, \mathbf{v}')$ plane (see formula (6)) from the relativistic electron scattered at the angle $\vartheta_e$. The charters near the curves show the values of parameter $\beta = \gamma \vartheta_e$.

For small values of the scattering angle ($\beta \ll 1$)

$$\frac{d^2 E}{d\omega\, do} \approx \frac{e^2 \gamma^2}{\pi^2} \beta^2 \frac{\left(1 - \alpha_x^2\right)^2}{\left(1 + \alpha_x^2\right)^4} \ . \tag{7}$$

This formula shows, that at $\beta \ll 1$ the maxima of the angular distribution of radiation are located at $\alpha_x = 0$, and that at $\alpha_x = \pm 1$ the spectral-angular density of electron radiation equals zero. The main body of the spectral density of radiation in this case is concentrated at angles $\alpha_x$ of the order of unity.

For large values of the scattering angle ($\beta \gg 1$) the angular distribution of radiation (6) has maxima at the angles $\alpha_x \approx 1$ and $\alpha_x \approx \beta - 1$, and equals zero at $\alpha_x \approx -1/\beta$ and $\alpha_x \approx \beta + 1/\beta$ (see Fig. 1). The formula (6) also shows that the angular density of radiation decreases rapidly at the angles $\alpha_x \leq -1$ and $\alpha_x \geq \beta + 1$, and in the range of angles $1 \leq \alpha_x \leq \beta$ the angular density of radiation has comparable values in a rather broad interval of scattering angles $\beta$. In particular, for $\beta = 10$ the minimum at $\alpha_x = \beta/2$ is only 50 % lower than the maximum. It means that at $\beta \gg 1$ the main body of the spectral density of radiation by the electron is concentrated in the range of angles $0 \leq \alpha_x \leq \beta$.

For $\beta \gg 1$, according to (5), the angular density of radiation in the directions close to the initial velocity of the particle is determined by the expression

$$\frac{d^2E}{d\omega do} \approx \frac{e^2\gamma^2}{\pi^2}\frac{\alpha^2}{(1+\alpha^2)^2}, \quad \alpha \ll \beta. \qquad (8)$$

In this case the angular density of radiation does not depend on the scattering angle of a particle.

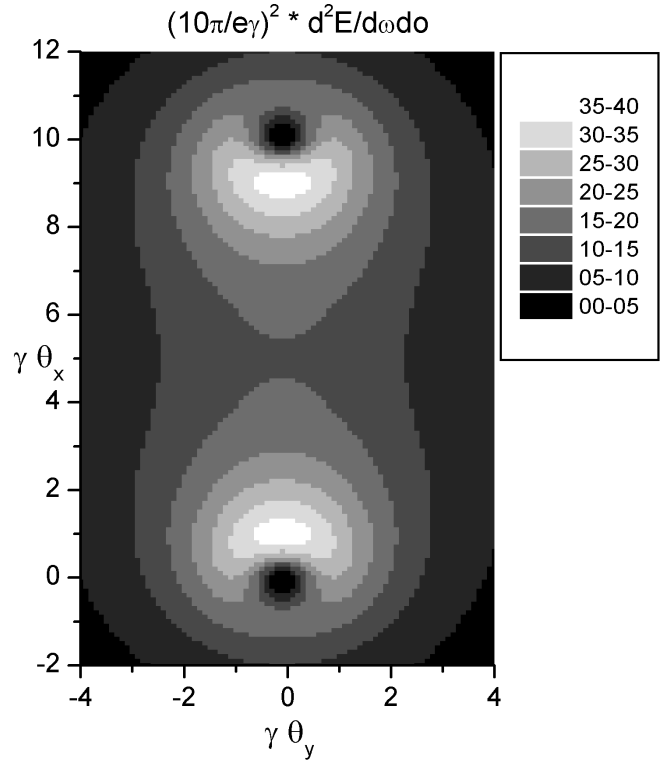

Fig. 2. The two-dimensional angular distribution of radiation by the relativistic electron scattered at the angle $\vartheta_e = 10\,\gamma^{-1}$.

The two-dimensional picture of the angular distribution of radiation by a relativistic electron scattered at the angle $\vartheta_e = 10\,\gamma^{-1}$ is given in Fig. 2. This picture shows a nontrivial angular distribution of essentially non-dipole electron radiation ($\beta \gg 1$) with sharp maxima and deep minima near the initial and final directions of the electron motion.

### 3. Multiple scattering effect

If we are interested in the angular distribution of radiation from an electron beam passing through a thin target, then the formula (5) is necessary for averaging over the scattering angles of the particles in matter. If the distribution function of the scattered particles $f(\vec{\vartheta}_e)$ is known, then the average value of spectral-angular density of radiation will be determined by the expression

$$\left\langle \frac{d^2E}{d\omega do} \right\rangle = \int d\vec{\vartheta}_e f(\vec{\vartheta}_e) \frac{d^2E}{d\omega do}. \qquad (9)$$

Note that formula (9) is applicable to any targets. It is only required that the target thickness is small in comparison with the coherence length of radiation. The different characters of the scatterer will be exhibited only by the definite kinds of distribution function $f(\vec{\vartheta}_e)$.

The particle distribution $f(\vec{\vartheta}_e)$ over the scattering angles for an amorphous target is determined by the Bethe-Molière function [8]. The multiple scattering effect on the spectral-angular distribution of the radiation by relativistic electrons in a thin amorphous target was studied in [7]. It was shown that even after the averaging procedure the minimum in the angular distribution of the emitted gamma-quanta is still observed in the initial direction of the electron beam (see Fig. 2 of the [7]).

When a beam of relativistic electrons passes through a crystal at a small angle to one of crystallographic axes there takes place a significant orientation effect in electron scattering, exhibited as a characteristic annular angular distributions of the particles outgoing from the crystal ("doughnut scattering") [10]. In this case the magnitude of the root-mean-square scattering angle of the electrons can exceed substantially (by several times) the corresponding parameter for the electron scattering in the amorphous target of the same thickness [5], and the smaller is the target thickness, the greater is this difference. This shows the existence of a coherent effect in electron scattering when the electron sequentially collides with the lattice atoms located along a given crystallographic axis.

Generally the dynamics of a relativistic particle beam in an aligned crystal is rather complicated, since various fractions of a beam are involved in various regimes of motion: finite and infinite, regular and chaotic, with transitions between them. The analytical description of the particle dynamics can be conducted only in some limiting cases. Thus, for example, the theory of multiple scattering of relativistic charged particles on atomic strings of a crystal, based on the continuous string approximation, describes the coherent azimuthal scattering of above-barrier electrons ("doughnut scattering" effect) [5,10]. However, this theory does not describe transitions of particles between two different fractions of the electron beam in the crystal, since the continuous string approximation does not take into account incoherent scattering. It is possible to take incoherent scattering into account by analytical methods [11] only in the case of rather large incident angles $\psi \gg \psi_L$, where $\psi_L$ is the Lindhard angle [10]. At the same time, as it was already mentioned, the orientation effects in scattering and radiation of a relativistic electron beam passing through a crystal are mostly manifest in the range of angles $\psi < \psi_L$. Therefore, for the quantitative description of these effects, a computer simulation of the passing of an electron beam through an aligned crystal appears to be the most adequate.

## 4. Computer simulation

With the purpose of a quantitative analysis of the multiple scattering effect on coherent radiation of relativistic electron in a thin crystal, we performed a computer simulation on the basis of the Monte-Carlo method. We used here the binary collisions model of the electron interactions with the atoms of a crystalline lattice. Such an approach allows to take into account both the coherent scattering of fast electrons on the atomic strings of the crystal and incoherent scattering of the electrons connected with the thermal fluctuations of the atom positions in the lattice and with the electronic subsystem of the crystal [12]. (Note that the incoherent scattering on the fluctuating part of the crystal potential is thus treated classically. For low Z, it maybe preferable to use quantum scattering. The result is not expected to be very different, however.) Rather small thickness of the crystal under consideration here (*T=10-100 μm*) allows to gather a sufficient statistics of events (*N = 10000*) during an acceptable period of time.

Figure 3 **a** represents the angular distribution of 10 GeV electrons scattered by a 10 μm silicon monocrystal when the electron beam is incident on the crystal at the Lindhard angle $\psi = \psi_L$ to the axis <111>. This picture is a regular angular distribution known as "doughnut scattering". Figure 3 **b** shows the angular distribution of gamma-quanta emitted by the electrons. One can see that the shape of the angular distribution of radiation is rather different from the one of the scattered electron beam. In particular, one can observe a depletion of the photon angular distribution at $\gamma\theta_y \sim 0$ and $\psi$ belonging to the interval $[\psi_L, \psi_L + \gamma^{-1}]$. This depletion is related to the left dip of Fig. 1. This feature can appear thanks to the superposition of two different effects: coherent electron scattering by the crystal rows ("doughnut scattering") and radiation suppression due to the multiple scattering effect (similar to the LPM effect in an amorphous target).

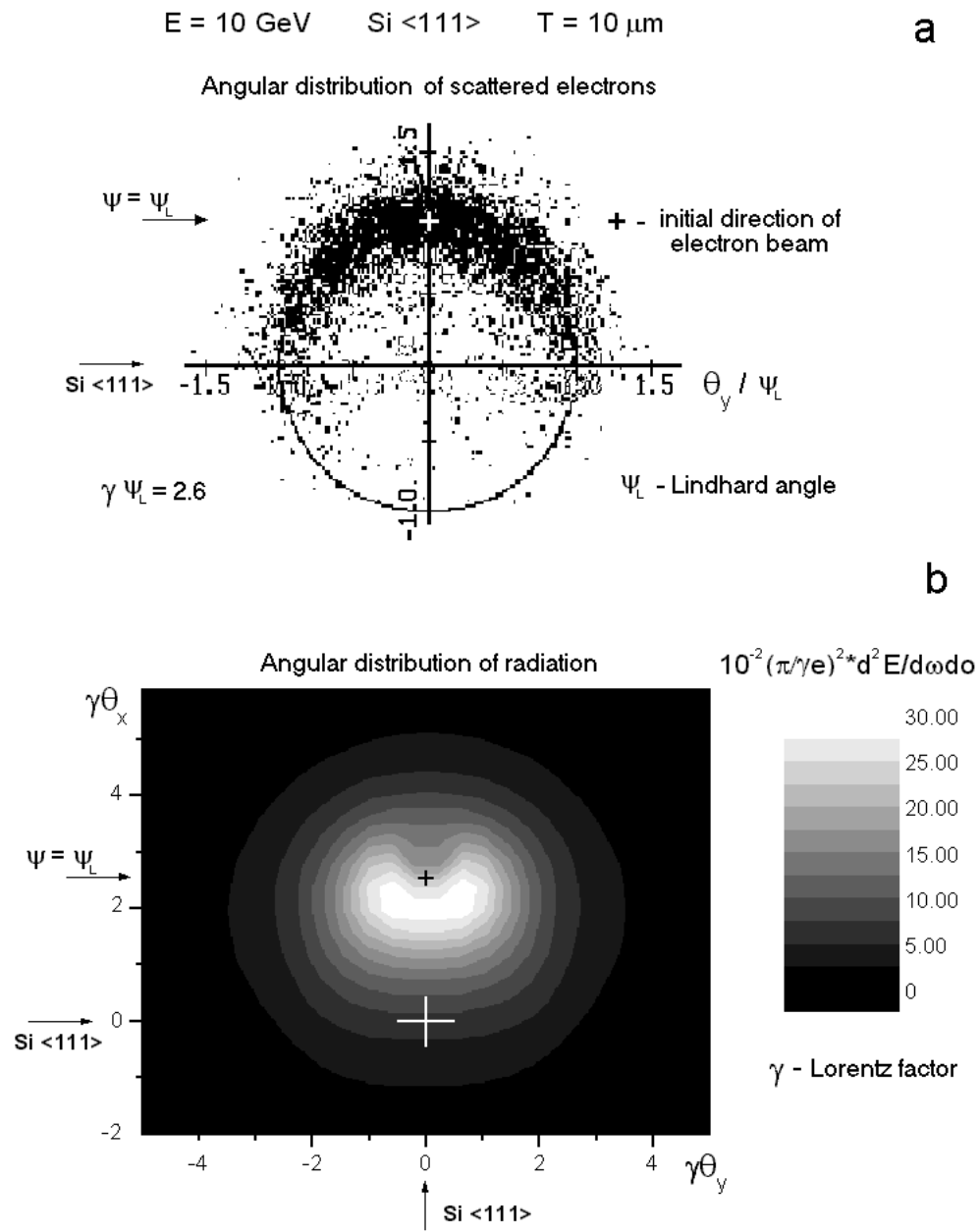

Fig. 3. The angular distribution of 10 GeV electrons scattered by a 10 μm silicon monocrystal when the electron beam is incident on the crystal at the Lindhard angle $\psi = \psi_L$ to the axis <111> (a) and the angular distribution of gamma-quanta emitted by these electrons (b).

Figure 4 represents the evolution of the angular distribution of emitted gamma-quanta with the change of the incident angle ψ from zero (a) to $4\psi_L$ (d). This figure demonstrates a complicated orientation behavior of the angular distribution of radiation.

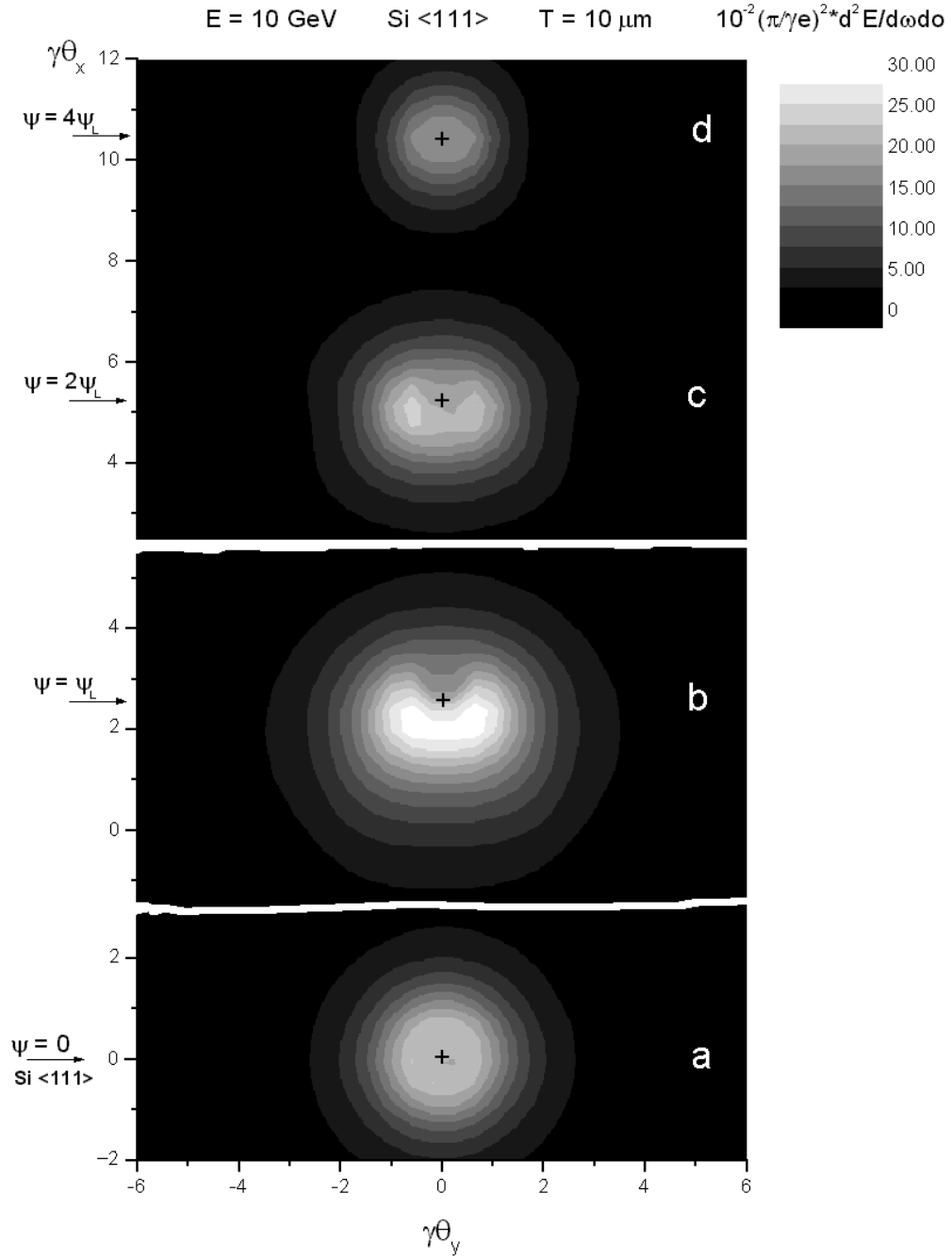

Fig. 4. The angular distributions of emitted gamma-quanta for the different values of the incident angle $\psi$ : (a) $\psi = 0$; (b) $\psi = \psi_L$; (c) $\psi = 2\psi_L$ and (d) $\psi = 4\psi_L$.

## 5. Conclusion

The present investigation shows a strong effect of the multiple scattering of relativistic electrons in a crystal on the angular distributions of their radiations in a relatively low energy region of emitted gamma-quanta.

For the experimental observation of the effect described above a high angular resolution (better than $\gamma^{-1}$) of the gamma-detector is needed, as well as a small (less than $\gamma^{-1}$) divergence of the electron beam. This effect must be taken into account when studying spectral-angular distributions of radiation by relativistic electrons in a crystal, especially when using a low divergence beam and a strongly collimated photon beam of about $\gamma^{-1}$.

## 6. Acknowledgments

The work is partly supported by the RFFR project # 03-02-16263 and by the STCU project # 1746. One of the authors S.P.F. also expresses gratitude to the staff of the Lyon Institute of Nuclear Physics for their hospitality and support in realization of this work.